\documentclass{article}

\usepackage{hyperref}

\usepackage{amsmath}
\usepackage{graphicx}
\usepackage{float}
\usepackage{caption}
\usepackage{subcaption}
\usepackage{amssymb}
\usepackage[toc,page]{appendix}
\usepackage{amsthm}

\newtheorem{thrm1}{Theorem}

\begin{document}

\title{A New Method for Triangular Mesh Generation}
\date{}

\maketitle

\begin{center}
\author{G. Liao \footnotemark[1] \footnotetext[1]{Department of Mathematics, University of Texas at Arlington, Arlington, TX, US}}
\author{X. Chen \footnotemark[1]}
\author{X. Cai \footnotemark[2] \footnotetext[2]{Nanhua Power Institute, Zhuzhou, China}}
\author{B. Hildebrand \footnotemark[1]}
\author{D. Fleitas \footnotemark[3] \footnotetext[3]{Department of Mathematics, Dallas Baptist University, Dallas, TX, US}}
\end{center}

\begin{abstract}
Computational mathematics plays an increasingly important role in computational fluid dynamics (CFD). The aeronautics and aerospace research community is working on next generation of CFD capacity that is accurate, automatic, and fast. A key component of the next generation of CFD is a greatly enhanced capacity for mesh generation and adaptivity of the mesh according to solution and geometry. In this paper, we propose a new method that generates triangular meshes on domains of curved boundary. The method deforms a Cartesian mesh that covers the domain to generate a mesh with prescribed boundary nodes.  The deformation fields are generated by a system of divergence and curl equations which are solved effectively by the least square finite element method.
\end{abstract}

\section{Introduction}

Despite considerable success, mesh generation and adaptation remain a challenging task for numerical simulation on complex geometries. A NASA sponsored study entitled CFD Vision 2030 states that “Mesh generation and adaptivity continue to be significant bottlenecks in the CFD workflow, and very little government investment has been targeted in these areas. As more capable HPC hardware enables higher resolution simulations, fast, reliable mesh generation and adaptivity will become more problematic. Additionally, adaptive mesh techniques offer great potential, but have not seen widespread use due to issues related to software complexity, inadequate error estimation capabilities, and complex geometries.” 

In this paper, we propose an innovative approach to the generation and adaptivity of triangular meshes that overcomes problems of current techniques.

We demonstrate the proposed method on a domain $D$, which resembles the right half of the curved domain in Figures 8.1 of the well-known textbook [George 1991]\cite{george}. In Figures 8.2, the so-called superposition-deformation method is described and demonstrated. This method, investigated by [Yerri et al. 1984] \cite{Yerri}, [Cheng et al. 1988]\cite{cheng} and [Shephard et al. 1988] \cite{shephard}, constructs a mesh of a curved domain with prescribed boundary nodes. A modified version is described in detail which uses quadtree-based partitioning of the squares near the boundary to better approximate the geometry of the boundary. After the squares near the boundary are sufficiently refined, all squares that intersect the domain are split into triangles. The boundary squares are treated based on a careful classification of the boundary patterns. The interior triangles are smoothed by iterations based on barycentric coordinates of the vertices. The final mesh generated by the modified method is shown in Figure 8.8. 
It is pointed out in [George 1991] \cite{george} that (1) Treatment of squares near the boundary is a difficulty of the method; (2) The generated mesh respects the initial contour; but, in general, it may contain a slightly different boundary discretization in that a given edge being the union of smaller edges. In fact, visual inspection of the mesh in Figure 8.8 reveals poor quality of the mesh near the boundary. In contrast, our method deforms a Cartesian mesh that covers the domain to generate a mesh on the domain using only the prescribed boundary nodes.  The deformation fields are generated by a system of divergence and curl equations which are solved effectively by the least squares finite element method. The deformation field preserves the Jacobian determinant and thus the method prevents element inversion. Moreover, the implementation of our method is based on differential equations, which do not rely on treatment of boundary squares on case-by-case bases. The resulted mesh uses the initial edges on the boundary. Thus the initial description of the geometry is preserved. In the next section, we use an example to demonstrate our method. In the following sections, the mathematical algorithm of the deformation method and the least squares finite element method used in the method are further discussed.

\section{An Example}
A uniform Cartesian mesh on the background domain $[1, 9]$ by $[1, 12]$ is shown in Figure \ref{fig:contour} below. The domain $D$ is the interior of the contour described by points $\# 1$ to $\# 18$ on the curved boundary, and by the nodes $(1, 5), (1, 6), (1, 7), (1, 8),$ $(1, 9)$, and $(1, 10)$ on the vertical boundary, see Figure \ref{fig:contour}. We select a set of 18 nodes on the Cartesian mesh that are close to the boundary of domain $D$, and move them to the points $\# 1$ through $\# 18$, respectively (see Appendix).  For instance, by our method, node $P_1(1, 4)$ is moved to $\# 1$; $P_2(1, 3)$ to $\# 2$; $P_3(2, 2)$ to point $\# 3$; $P_4(3, 2)$ to $\# 4$; node $P_5(4, 2)$ to $\# 5$; $P_6(4, 3)$ to $\# 6$;  $P_7(5, 3)$ to $\# 7$; $P_8(6, 4)$ to $\# 8$; node $P_9(7, 4)$ to point $\# 9$; node $P_{10}(8, 5)$ to $\# 10$; $P_{11}(7,6)$ to $\#11$; $P_{12}(6, 7)$ to $\# 12$; $P_{13}(5, 8)$ to $\# 13$; $P_{14}(4, 8)$ to $\# 14$; $P_{15}(4, 9)$ to $\# 15$; $P_{16}(3, 9)$ to $\# 16$; $P_{17}(2, 10)$ to $\# 17$; $P_{18}(1, 11)$ to $\# 18$, etc. Correct movements of all other nodes are enabled by solving a divergence-curl system with suitable Dirichlet conditions on these 18 nodes as if they were boundary points. The selected background nodes are the small
dots in Figure b, which are moved to the prescribed boundary nodes $\#1 - \#18$, respectively. In Figure \ref{fig:drawnContour}, the prescribed boundary of domain $D$ is colored in blue. Finally, a triangular mesh is generated by connecting suitable pairs of the nodes on domain $D$, see Figure \ref{fig:traingles}. 

\begin{figure}[H]
	\centering
		\begin{subfigure}[b]{0.5\textwidth}
			\includegraphics[scale=.3]{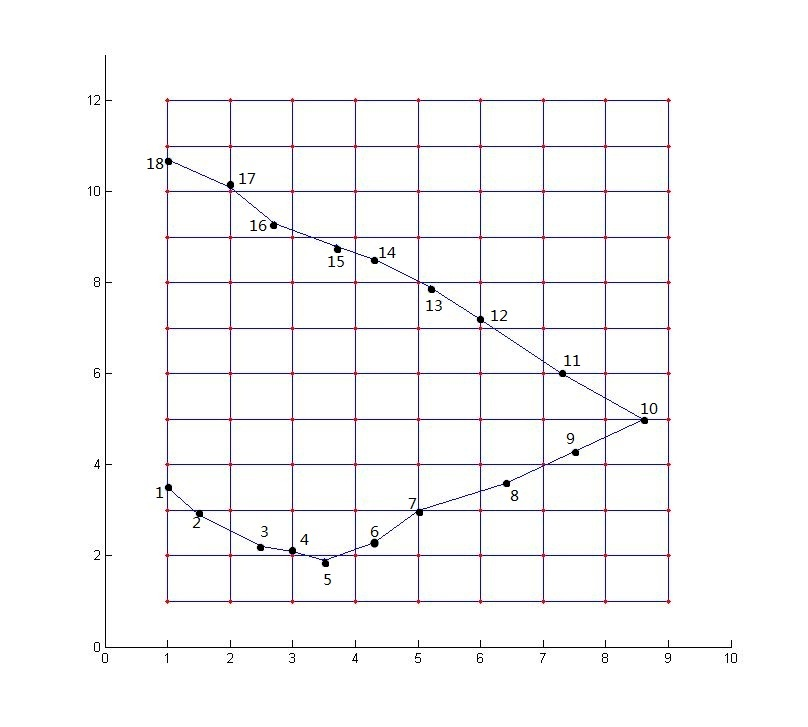}
			\caption{}
			\label{fig:contour}
		\end{subfigure}
		\begin{subfigure}[b]{0.45\textwidth}
			\includegraphics[scale=.5]{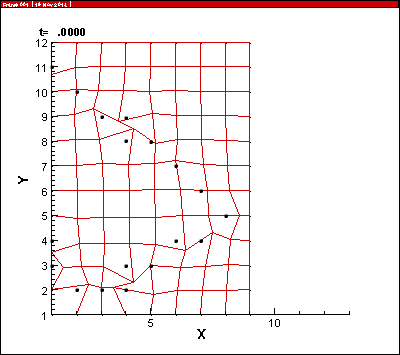}
			\caption{}
			\label{fig:moved}
		\end{subfigure}
\caption{}
\end{figure}

\begin{figure}[H]
	\centering
		\begin{subfigure}[b]{0.45\textwidth}
			\includegraphics[scale=.85]{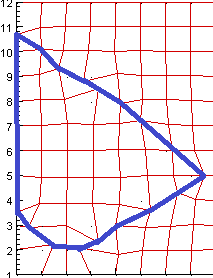}
			\caption{}
			\label{fig:drawnContour}
		\end{subfigure}
		\begin{subfigure}[b]{0.4\textwidth}
			\includegraphics[scale=.5]{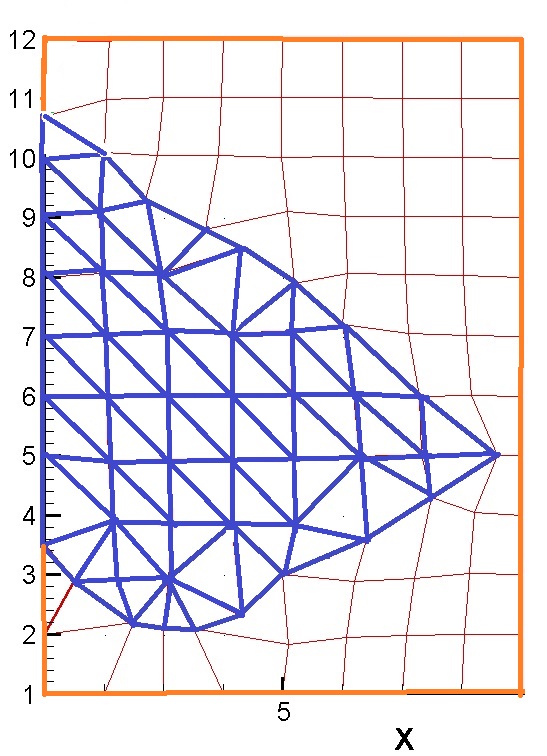}
			\caption{}
			\label{fig:traingles}
		\end{subfigure}
	\caption{}
\end{figure}

\section{The Deformation Method}
We now describe the original deformation method in differential geometry, which we adopted to generate non-folding meshes. 

\subsection{The Deformation Method of Jurgen Moser in Differential Geometry}
The method we propose has its origin in differential geometry. It was used in [Moser 1965] \cite{moser} for a study of volume elements in Riemannian manifolds and was modified for domains in $\mathbb{R}^d$ ($d$ is the dimension of the space) with boundary in [Dacorogna et al. 1990] \cite{dacorogna}. They showed how to construct a diffeomorphism (i.e. a smooth and invertible deformation) from a 2- or 3-D domain $\Omega_1$ to $\Omega_2$ with prescribed Jacobian determinant. The boundary nodes of $\Omega_1$ are matched to suitable boundary nodes of $\Omega_2$ in a one-to-one correspondence. For a prescribed positive size function $f(x, y, z) > 0$, a method based on Poisson equations is devised which determines a suitable velocity vector field $V(x, y, z, t)$. Each interior point of the domain is moved by $V$ from the artificial time $t = 0$ to $t = 1$. At each time $t$, a deformation $T(x, y, z, t)$ is constructed by the motion. Moser and Dacorogna proved that the final deformation, at $t = 1$, has the prescribed Jacobian determinant, i.e. its Jacobian determinant $J(T(x, y, z, 1) = f(x, y, z)$. Since $f(x, y, z)$ is positive, $T(x, y, z, 1)$ has the desirable property of having positive Jacobian determinant everywhere in the domain.

\subsection{Adaptive Non-folding Mesh Generation and Adaptation on Moving Domains}
We now present our deformation method on moving domains. To generate a mesh on a domain $\Omega_2$, we deform an initial mesh on $\Omega_1$ by a discrete approximation of a smooth and invertible deformation $\phi$ from $\Omega_1$ to $\Omega_2$. In fact, a family of adaptive non-folding meshes can be numerically generated if a positive, “time” dependent size function $f(\chi,t) > 0$ is specified, where $\chi = (x, y)$ in 2D, $t \ge 0$ is a parameter (it could be the real time in unsteady field simulation).

In this subsection, we describe the deformation method for generation of a family of deformations according to a prescribed (“time” dependent) size function $f(\chi,t) > 0$. We will explain how to construct higher order meshes below. 

The problem to solve is the following: Given a size function $f(\chi,t) > 0$, find deformations $\phi (x,y,t)$ such that the Jacobian determinant of the deformations are equal to the given size function $f(\chi,t)$.This problem is solved in the following two steps.

Step 1: Find a vector field $U$ by solving the divergence-curl system:

\begin{eqnarray}
\nonumber \mbox{div}_\phi U(\phi,t) & = & \frac{\partial}{\partial t} \frac{1}{f(\phi,t)} \\
\nonumber & = & \frac{\partial}{\partial t} g(\phi,t) \\
\nonumber \mbox{curl}_\phi U(\phi,t) & = & 0
\end{eqnarray}

\noindent on $\Omega_1$, with the Dirichlet condition on moving portion of the boundary, and the Neumann condition on fixed boundary. For later use we have introduced $g=\frac{1}{f}$.

Step 2: Define the velocity vector field by $V = f U$ and solve the ordinary differential equation for the transformation:

\begin{eqnarray}
\nonumber \frac{\partial \phi (x,t)}{\partial t} & = & f(\phi,t) U(\phi,t) \\
\nonumber & = & V(\phi,t)
\end{eqnarray}
 
\noindent with the initial condition $\phi (x,y,0) = (x,y)$. From [Cai et al. 2004]\cite{cai} we have the following theorem.

\begin{thrm1}
The deformation constructed by the above steps satisfies, $\forall t > 0, \; \emph{det} \nabla \phi(x,t) = f(\phi,t)$.
\end{thrm1}

Indeed, defining $H = \frac{J(\phi(x,y,t))}{f(\phi(x,y,t))}$, we can show $ \frac{\mbox{d}H}{\mbox{d} t} = 0$ directly based on the two steps, which then implies that $H$ is constant. For the details, we refer to the dissertation by Dion Fleitas.

\section{The Numerical Implementation}

We use the example in section 2 to describe the numerical implementation of the deformation method.

We denote the background domain by $\Omega(0)$. We will deform the initial Cartesian mesh on $\Omega(0)$ by a suitable velocity vector from $t=0$ to $t=1$, and denote the deforming domain by $\Omega(1)$. In the example, $\Omega(t) = \Omega(0)$. Suppose we want $P_i$ on $\Omega(0)$ to be moved to $Q_i$ on $\Omega(1)$. For instance, in the example, we want node $P_1 = (1, 4)$ to be moved to $Q_1 = \# 1$; $P_2 = (1, 2)$ to $Q_2 = \# 3$; $..., P_{18} = (1, 11)$ to $Q_{18} = \# 18$, etc. A pseudocode for selecting the nodes $P_i$ as a general case is provided in the Appendix.

In order to define a correct velocity vector field $V$, we first find a vector field $U(\phi,t)$ by solving the divergence-curl system on the deforming domain $\Omega(t)$ at each time $t = \frac{k}{n}$, where $n$ is the number of time steps for solving the ordinary differetial equation (ODE) from $t=0$ to $t=1$, $k = 1, 2, ..., n-1$. In the example, we take $n = 10$. The following equations are solved by the least squares finite element method (as described in [Cai et al. 2004] \cite{cai} and [Liao et al. 2009] \cite{liao}) at each fixed time $t = \frac{k}{n}: $

\begin{eqnarray}
\nonumber \mbox{div}_\phi U(\phi,t) & = & \frac{\partial}{\partial t} \frac{1}{f(\phi,t)} \\
\nonumber & = & \frac{\partial}{\partial t} g(\phi,t) \\
\nonumber \mbox{curl} _\phi U(\phi,t) & = & 0
\end{eqnarray}

\noindent with the Dirichlet condition $U = \frac{P_i Q_i}{n \cdot f}, i = 1, 2, 3, ..., 18$ in the example, and $U = 0$ at all other boundary nodes of the background domain, where $P_i Q_i$ is a vector from $P_i$ to $Q_i$. Solving the ODE:

After determination of $U$ at $t = \frac{k}{n}, k = 1, 2, ..., n-1$, we define the velocity vector field by $V(\phi,t) = f(\phi,t) U(\phi,t)$ and solve the ODE for the transformation $\phi(x,t)$ from $t = 0$ to $t = 1$, in $n$ time steps:

\begin{eqnarray}
\nonumber \frac{\partial \phi(x,t)}{\partial t} & = & f(\phi,t) U(\phi,t)
\end{eqnarray}

\noindent with the condition $\phi (x,0) = x$.

The deformed mesh on the background domain is now divided into two parts: the exterior part is disregarded as shown in Figure \ref{fig:drawnContour}; the interior part is reconnected to be a triangular mesh in a straight forward manner as shown in Figure \ref{fig:traingles}. If an exterior mesh is desired, then we disregard the interior mesh.

\appendix

\section{Selection of Background Mesh Nodes $P_i$'s}
\label{app:A}

\noindent We can lay a background mesh of step size $h$ over a contour such that each cell of the mesh contains only one contour point (provided the mesh won't be too dense). Then we need only to find the shortest distance between the contour point of interest and the four points that make up the cell containing that contour point to avoid searching for the shortest distance between the contour point and all mesh points. To do this we search for the minimium distance from one contour point to another, say, $d_c$, and then let $h = \frac{0.9 \cdot d_c}{\sqrt{2}}$ be the step size of the mesh. Let $y_k = (t_k, s_k)$ be the $k$th point on the contour and $x_{i,j}$ a mesh point. Let $B_{x_{i,j}}$ and $B_{y_k}$ be elements of Boolean matrices determining if a mesh point, $x_{i,j}$, has been moved to a contour point or if a contour point, $y_k$, has had a mesh point moved to it, respectively. Then the following pseudo-code selects a set of background mesh points that will be moved to the corresponding contour points.
\newline \newline \noindent \textbf{for} $ ( k = 1 $ to $n ) $
\newline \indent $ t = \frac{t_k}{h}$;
\newline \indent $ s = \frac{s_k}{h}$;
\newline \indent $ i = \lfloor t \rfloor $;
\newline \indent $ j = \lfloor s \rfloor $;
\newline \indent $ d_1 = distance(x_{i,j}, y_k) $;
\newline \indent $ d_2 = distance(x_{i+1,j}, y_k) $;
\newline \indent $ d_3 = distance(x_{i,j+1}, y_k) $;
\newline \indent $ d_4 = distance(x_{i+1,j+1}, y_k) $;
\newline \indent $ d_{min} = min( d_1, d_2, d_3, d_4 ) $;
\newline \indent \textbf{if} $ ( d_{min} == d_1 ) $ 
\newline \indent \indent \textbf{if} $ B_{x_{i,j}} == false $
\newline \indent \indent \indent $x_{i,j}$ = $y_k$;
\newline \indent \indent \indent $B_{x_{i,j}} = true$;
\newline \indent \indent \indent $B_{y_k} = true$;
\newline \indent \indent \textbf{else if} $ (B_{x_{i,j}} == true )$
\newline \indent \indent \indent $ d_{min} = min( d_2, d_3, d_4 ) $;
\newline \indent \textbf{if} $ ( d_{min} == d_2 ) $ 
\newline \indent \indent \textbf{if} $ B_{x_{i+1,j}} == false $
\newline \indent \indent \indent $x_{i+1,j}$ = $y_k$;
\newline \indent \indent \indent $B_{x_{i+1,j}} = true$;
\newline \indent \indent \indent $B_{y_k} = true$;
\newline \indent \indent \textbf{else if} $ (B_{x_{i+1,j}} == true )$
\newline \indent \indent \indent $ d_{min} = min(  d_3, d_4 ) $;
\newline \indent \textbf{if} $ ( d_{min} == d_3 ) $ 
\newline \indent \indent \textbf{if} $ B_{x_{i,j+1}} == false $
\newline \indent \indent \indent $x_{i,j+1}$ = $y_k$;
\newline \indent \indent \indent $B_{x_{i,j+1}} = true$;
\newline \indent \indent \indent $B_{y_k} = true$;
\newline \indent \indent \textbf{else if} $ (B_{x_{i,j+1}} == true )$
\newline \indent \indent \indent $ d_{min} = min( d_4 ) $;
\newline \indent \textbf{if} $ ( d_{min} == d_4 ) $ 
\newline \indent \indent \textbf{if} $ B_{x_{i+1,j+1}} == false $
\newline \indent \indent \indent $x_{i+1,j+1}$ = $y_k$;
\newline \indent \indent \indent $B_{x_{i+1,j+1}} = true$;
\newline \indent \indent \indent $B_{y_k} = true$;
\newline \noindent \textbf{end}

\bibliography{DeformingCartesianMeshes}

\end{document}